# Computational study of exciton generation in suspended carbon nanotube transistors

*Siyuranga O. Koswatta,[†,*] Vasili Perebeinos,[‡] Mark S. Lundstrom,[†] and Phaedon Avouris[‡]*

† 465 Northwestern Ave, Electrical and Computer Engineering, Purdue University, West Lafayette, IN 47906, USA

‡ IBM T. J. Watson Research Center, Yorktown Heights, NY 10598, USA

* Corresponding Author – koswatta@purdue.edu

Abstract - Optical emission from carbon nanotube transistors (CNTFETs) has recently attracted significant attention due to its potential applications. In this paper, we use a self-consistent numerical solution of the Boltzmann transport equation in the presence of both phonon and exciton scattering to present a detailed study of the operation of a partially suspended CNTFET light emitter, which has been discussed in a recent experiment. We determine the energy distribution of hot carriers in the CNTFET, and, as reported in the experiment, observe localized generation of excitons near the trench-substrate junction and an exponential increase in emission intensity with a linear increase in current versus gate voltage. We further provide detailed insight into device operation, and propose optimization schemes for efficient exciton generation; a deeper trench increases the generation efficiency, and use of high-k substrate oxides could lead to even larger enhancements.

In recent years, rapid progress in research on carbon nanotube (CNT) electronic and optoelectronic devices has occurred.[1,2] High performance CNT transistors operating close to the ballistic limit have been experimentally realized.[3-5] The demonstration of electroluminescence in CNTs has opened up new possibilities for optoelectronic applications as well.[6-9] The importance of nanoscale physical phenomena in governing the operation of CNT devices has been underscored in a multitude of experiments.[1,2] Optical emission in earlier CNT devices [6,7] was attributed to free carrier recombination under ambipolar transport, which is similar to the operation of conventional light emitting diodes (LEDs). In this case, a Schottky-barrier CNT field-effect transistor (CNTFET) is operated at high source-drain bias ($V_{DS}$) such that both electrons and holes are injected into the channel simultaneously from the two contacts, respectively.[6,7] These free carriers radiatively recombine inside the channel region resulting in optical emission. Device operation was well modeled by drift-diffusion treatment of carrier transport and free carrier recombination.[10-13] A recent experiment,[8] however, using a partially suspended CNTFET as shown in Fig. 1(a) demonstrated an enhancement of optical emission intensities of 100-1000x than that in the earlier work. The emission process was attributed to *exciton generation under unipolar transport conditions*.[8]

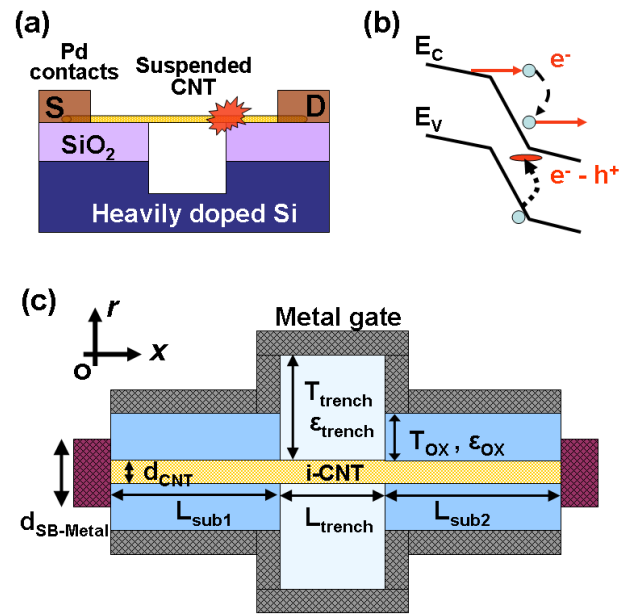

Fig 1. (a) A schematic of the partially suspended CNTFET geometry reported in the experiment.[8] The trench region is fabricated by reactive ion etching. Localized optical emission is observed at the trench-substrate junction. (b) Band diagram near the trench-substrate junction at the drain end (shown for $V_{GS} > 0$). Electrons that are injected from the source can gain enough kinetic energy at this junction for the impact excitation of excitons. (c) Model device structure used in this study with a cylindrically symmetric geometry with the oxide ($\varepsilon_{OX}$) and the trench ($\varepsilon_{trench} = 1$) regions wrapping around the CNT. Source/drain metal Schottky contacts have been used. The gated channel region ($L_{sub1}+L_{trench}+L_{sub2}$) is undoped.

It has already been well established that the optical processes in CNTs are dominated by excitonic effects [14, 15] due to the increased Coulomb interactions in these quasi 1-dimensional structures.[16-20] In semiconductor CNTs,[21] the exciton binding energies, and the ionization fields that correspond to the stability of bound electron-hole pairs under the influence of an electric field, are nearly two orders of magnitude stronger than that in conventional, bulk semiconductors.[22] Therefore, efficient impact excitation of stable excitons is expected in CNT devices under suitable operating conditions.[23] In the case of the partially suspended CNTFET reported in Ref. 8, as shown in Fig. 1(b), there exists a high electric-field region at the trench-substrate junction near the drain contact (shown for positive gate bias, $V_{GS} > 0$). This arises due to the difference in electrostatic coupling to the back-gate in the trench and the substrate regions, respectively. Electrons injected from the source can gain enough kinetic energy in this high-field region to generate excitons by impact excitation. The subsequent radiative recombination of these excitons is expected to lead to localized optical emission at the trench-substrate junction seen in Ref. 8. At this point it is important to differentiate the above mechanism from thermal light emission reported in suspended *metallic* CNTs.[24-26] In the latter case the optical emission is seen in the *middle* of the trench region where the local temperature is the highest.[24, 27]

In this paper we present a detailed numerical investigation of the partially suspended CNTFET reported in Ref. 8. Our model treats inelastic effects on carrier transport due to both optical phonon (OP) scattering and the excitonic process. It should be noted that the prior work [10-13] on device modeling of optical emission in CNTFETs operating in the diffusive limit did not consider the latter process. The model device structure (which represents the more complex 3-dimensional geometry of the experiment) is shown in Fig. 1(c). It has cylindrical symmetry with gate oxide and trench regions that wrap around the CNT. A (25,0) zigzag CNT is used unless specified otherwise. As in the experiment,[8] source/drain metal Schottky contacts were used in the simulation. (A preliminary study was also carried out for a CNT-MOSFET with doped source/drain contacts.[28]) The 1D Boltzmann transport equation (BTE) was numerically solved for electron transport in the lowest conduction band,[29, 30]

$$\left[\frac{\partial}{\partial t} + v_k \cdot \frac{\partial}{\partial x} + \frac{(-q\xi_x)}{\hbar} \cdot \frac{\partial}{\partial k}\right] f(x,k) = S^{in} - S^{out} \quad (1)$$

where, $v_k$ is the band velocity, $\xi_x$ is the position dependent electric field, $f(x,k)$ is the phase-space distribution function, and $S^{in/out}$ are the in/out-scattering rates described below. In calculating $v_k$, CNT energy dispersion of $E(k) = (3a_{CC}\gamma/2)\sqrt{k^2 + \Delta k^2} - (E_G/2)$ has been used where $a_{CC} = 0.142$nm is the carbon-carbon bond length, $\gamma = 3$eV is the hopping parameter, $\Delta k$ is the angular (confinement) momentum of the sub-band, and $E_G = 3a_{CC}\gamma\Delta k$ is the CNT bandgap.[31] A similar solution of the BTE has been used in Ref. 32 to investigate hot-carrier effects in conventional CNT-MOSFETs (with doped source/drain contacts and without the trench geometry). Transport calculations were performed self-consistently with electrostatics by solving the 2D Poisson equation using the finite difference method.

When solving (1), appropriate boundary conditions needs to be applied near the contacts in order to account for tunneling through the Schottky barriers (see Supporting Information for details).[12, 31] At position $x_i$ near the source-end for an electron with momentum $k_j > 0$ (i.e: source-injected electron) we impose the following boundary condition on the phase-space distribution function:

$$f(x_i, k_j) = \overline{T} \bullet f_{F-D}\left(E_C(x_i) + E(k_j) - E_{FS}\right)$$
$$+ \left[1 - \overline{T}\right] \bullet f(x_i, -k_j) \quad (2)$$

where $E_C(x_i)$ is the conduction band edge, $E(k_j)$ is the kinetic energy of the electron from the band edge, $E_{FS}$ is the source Fermi energy, and $f_{F-D}$ is the Fermi-Dirac function. In (2) $\overline{T}$ is the transmission probability at energy $\left(E_C(x_i) + E(k_j)\right)$ calculated using the Wentzel-Kramers-Brillouin (WKB) approximation.[31] A similar boundary condition is applied at the drain-end for the drain-injected electrons (with $k_j < 0$).

The in/out-scattering rates in (1), $S^{in/out}$, are determined microscopically, accounting for the Pauli blocking of scattering events,[29]

$$S^{in}(x,k) = \sum_{k'} f(x,k')[1 - f(x,k)]S(k',k) \quad (3)$$

$$S^{out}(x,k) = \sum_{k'} f(x,k)[1 - f(x,k')]S(k,k') \quad (4)$$

where $S(k,k')$ is the transition rate from state-$k$ to state-$k'$, determined from Fermi's golden rule;

$$S_{OP,exc}(k,k') = \frac{2\pi}{\hbar}|M_{OP,exc}|^2 \delta\left(E(k') - E(k) + \hbar\omega_{OP,exc}\right) \quad (5)$$

Here, both OP emission [30] and exciton emission [23] processes by the high energy carriers are treated separately where $|M_{OP,exc}|$ is the matrix element for the respective scattering mechanism. For OP scattering, the zone boundary optical mode (180meV) and longitudinal optical mode (190meV) that have the largest electron-phonon coupling in CNTs were treated.[30] In this work, excitons were assumed to have a constant energy dispersion ($\hbar\omega_{exc}$ = constant) with an impact excitation energy threshold of 430meV for the (25,0) CNT considered here.[21, 23] Due to the treatment of the in/out-scattering processes in the degenerate limit in (3)-(4), (1) becomes nonlinear. Therefore, it is solved iteratively for the steady-state distribution function using an explicit upwinding scheme for the discretization of the BTE operators.[30] A full account of the solution procedure is given in Ref. 30. Finally, the exciton generation profile along the CNT, $R_{exc}(x)$, is determined from,

$$R_{exc}(x) = \sum_k S_{exc}^{out}(x,k) \qquad (6)$$

where, $S_{exc}^{out}(x,k)$, given by (4), is the steady-state out-scattering rate due to exciton emission by the high energy carriers.

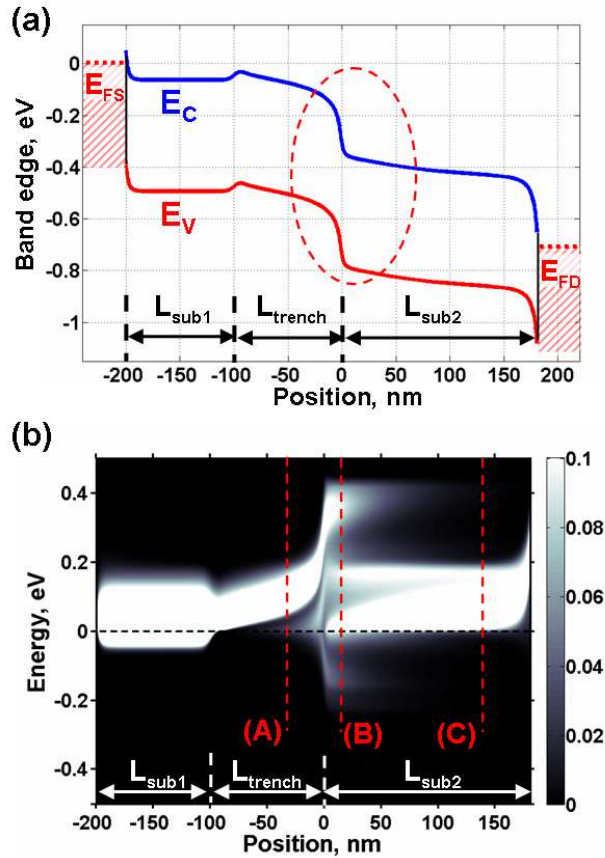

Fig. 2. (a) Band diagram for the (25,0) CNTFET at $V_{GS} = 0.8$V and $V_{DS} = 0.7$V calculated from self-consistent simulation. The device dimensions are, $T_{OX} = 10$nm (SiO$_2$, $\varepsilon_{OX} = 3.9$), $T_{trench} = 100$nm, $L_{sub1} = 100$nm, $L_{trench} = 100$nm, $L_{sub2} = 180$nm, and $d_{SB-Metal} = 10$nm (Schottky barrier height, $\varphi_{SB} = 50$meV). (b) Energy-position resolved electron distribution function for the above device. Note that positive (negative) energy represents positive (negative) going carries with $+k$ ($-k$) momentum. A significant hot-

electron population is seen at the trench-substrate junction near the drain-end (Line-(B)). See text for a detailed discussion.

A self-consistent energy band diagram under typical device operation is shown in Fig. 2(a). Device parameters used here are given in the caption. A characteristic Schottky barrier band profile is clearly seen near the contacts. Furthermore, as suggested in Ref. 8, a significant potential drop occurs at the trench-substrate junction near the drain-end (dashed circle). The band diagram observed here has also been inferred by recent experimental evidence.[33, 34] In the high-field region near the drain-end of the trench electrons can gain enough kinetic energy leading to a significant hot electron population. This can be clearly observed in the energy-position resolved distribution function for conduction band electrons shown in Fig. 2(b). In constructing Fig. 2(b) the momentum-energy conversion of the phase-space distribution function, $f(x,k)$, is accomplished using the dispersion relation, $E(k)$. The direction of electron momentum is given by the sign of the energy; i.e. $+k$ ($-k$) momentum depicted by positive (negative) energy. Inside the trench region, as seen in Fig. 2(a), the carriers are *gradually* accelerated under the moderate electric field. When they gain energy above the 180meV OP energy, they emit optical phonons and start to populate the states near the bottom of the band with $E \approx 0$. This is manifested by the appearance of low energy electron occupation near Line-(A) of Fig. 2(b).

The distribution function at Line-(B) situated just beyond the drain-end of the trench shows a significant population of hot carriers, and is elaborated on in Fig. 3. Here, it clearly shows the main mechanisms for carrier scattering near this region; OP and exciton emission. The exciton emission rate by hot carriers satisfying the energy threshold requirement is expected to be stronger (in the range of ~ $10^3$ ps$^{-1}$) [23] compared to energy loss through phonon scattering ($< 10^2$ ps$^{-1}$).[30] Therefore, efficient exciton emission, accompanied by OP emission, is observed in the evolution of the distribution function in this region (Fig. 3). By the time carriers reach Line-(C) of Fig. 2(b) all the hot electrons have been thermalized by emission of phonons/excitons and give rise to a stream of forward going carriers with energies delineated by the OP energy. Note that OP emission near the trench-substrate junction is expected to lead to Joule heat dissipation, especially in the high bias regime. However, the high thermal conductivity of CNTs and the effective coupling of CNT phonons to the substrate in this region would help reduce the local temperature. This is confirmed by both experimental [35] and device simulation studies.[30] We note that in metallic CNTs, where the OPs can efficiently decay to electron-hole pairs, thermal emission was observed, but from the suspended section of tubes (middle of tube), while it was absent for on-substrate tubes at comparable device currents.[24, 26] The non-thermal excitation of OPs, on the other hand, does lead to an enhancement of CNT light emission by impact excitation as discussed in Ref. 23. A more extensive analysis should include the self-heating effect and carrier multiplication effects at very high biases, but this is not expected to change the key conclusions of the paper.

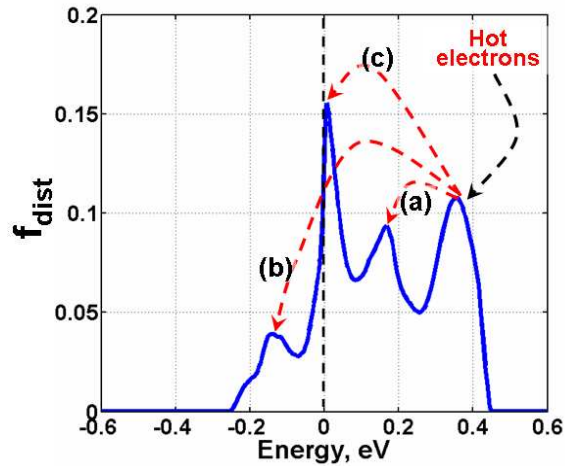

Fig. 3. The distribution function at Line-(B) of Fig. 2(b) ($x = 10$nm). As earlier, positive (negative) energy represents positive (negative) going carries with $+k$ ($-k$) momentum. The significant population of hot-electrons in this region can scatter to empty low-energy states by the following processes: (a) OP forward emission into the $+k_{final}$ state, (b) OP backward emission into the $-k_{final}$ state, or, (c) high-energy exciton emission that scatters electrons down to the states near the bottom of the band ($E \approx 0$).

The exciton generation profile along the CNT, i.e. $R_{exc}(x)$ from (6), is shown in Fig. 4(a). It clearly shows the localized nature of generation near the trench-substrate junction. Subsequent radiative recombination of these excitons leads to localized optical emission seen in the experiment.[8] Figure 4(a) also shows exciton generation near the drain contact. This is expected from the large electric field at that location as seen from Fig. 2(a). Optical emission near the drain, however, is expected to be largely quenched because the nanotube emission yield is lower on the oxide than from the suspended tube (possibly due to increased non-radiative decay channels), and near the metal contacts the excitons are quenched by energy transfer to plasmons and screening effects. Figure 4(b) shows that the highest generation rate near the trench-substrate junction is located to the right of the high-field region. This is easily understood by the energy threshold requirement for exciton emission by carriers accelerated under this field. Furthermore, it is expected to suppress the possibility of field-induced dissociation of excitons, and improve their radiative recombination. It is also observed that the exciton generation *cannot* be described by a local field-dependent generation rate as would be necessary for a drift-diffusion solution of this problem.

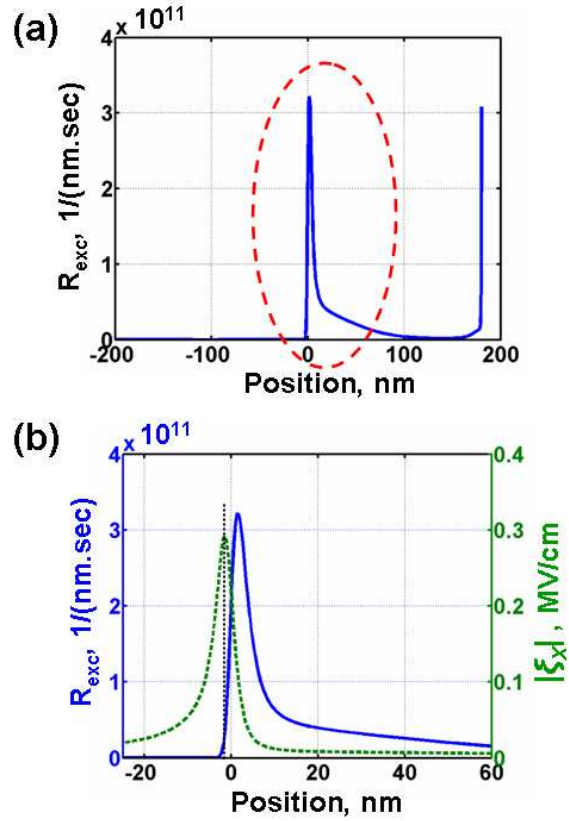

Fig . 4. (a) Steady-state exciton generation profile along the CNT for the device shown in Fig. 2. Localized exciton generation near the trench-substrate junction is clearly observed (dashed circle). The radiative recombination of these excitons leads to localized optical emission seen in the experiment.[8] Additional exciton generation near the drain contact is also observed. (b) The generation profile (solid) along with the electric field (dashed) near the trench-substrate junction. The highest generation region is located to the right of the high-field region.

Figure 5 shows the bias dependence of device characteristics; $I_{DS}$ vs. $V_{GS}$ and the exciton generation efficiency, $\eta_{exc} = (qN_{ex-tot}/I_{DS})$ vs. $V_{GS}$. In calculating the generation efficiency, $q$ is the electron charge, and $N_{ex-tot}$ is the total number of excitons generated per second near the trench-substrate junction. $N_{ex-tot}$ is calculated by integrating the spatially resolved generation rate, $R_{exc}(x)$, within the interval -50nm $\leq x \leq$ 120nm of Fig. 4(a). We first concentrate on the results for the $SiO_2$ substrate oxide with $T_{trench}$ = 10nm (solid circles). A linear increase in $I_{DS}$-$V_{GS}$ for above-threshold operation accompanied by an exponential increase in $\eta_{exc}$ vs. $V_{GS}$ is seen in this case. The latter relationship also leads to an exponential increase in the localized exciton generation rate, $N_{ex\text{-}tot}$. A key experimental result [8] was the observation of an exponential increase in light emission intensity vs. $V_{GS}$ with a linear increase of $I_{DS}$-$V_{GS}$; a relationship attributed to the impact excitation mechanism of the optical emission in these devices. Here, we confirm a similar relationship for the localized exciton generation rate that would lead to the experimentally observed device characteristics.

The effect of CNT diameter (bandgap) on exciton generation efficiency is explored in Fig. 5(b) by comparing the results for the (25,0) CNT to that of a (19,0) CNT device. For the latter, all device parameters are similar to that of the (25,0) CNT with SiO$_2$ substrate and $T_{trench}$ = 10nm (solid circles). We also assume a Schottky barrier height of $\varphi_{SB}$ = 50meV for carrier injection and the gate metal work function is chosen such that both devices have the same threshold voltage. Therefore, we obtain similar device currents ($I_{DS}$) for both the devices. The exciton generation efficiency for the (19,0) CNT, however, is observed to be significantly smaller (Fig. 5(b)). This is because of the larger electronic bandgap for the (19,0) CNT that increases the impact excitation energy threshold for exciton emission to 560meV. Similar behavior is also reported for the reduction of impact ionization rate with increasing energy threshold in CNT-MOSFET devices.[32] For smaller diameter tubes the device current itself could be reduced due to larger Schottky barriers for carrier injection.[36] This would further reduce the total exciton generation rate for smaller diameter CNTs.

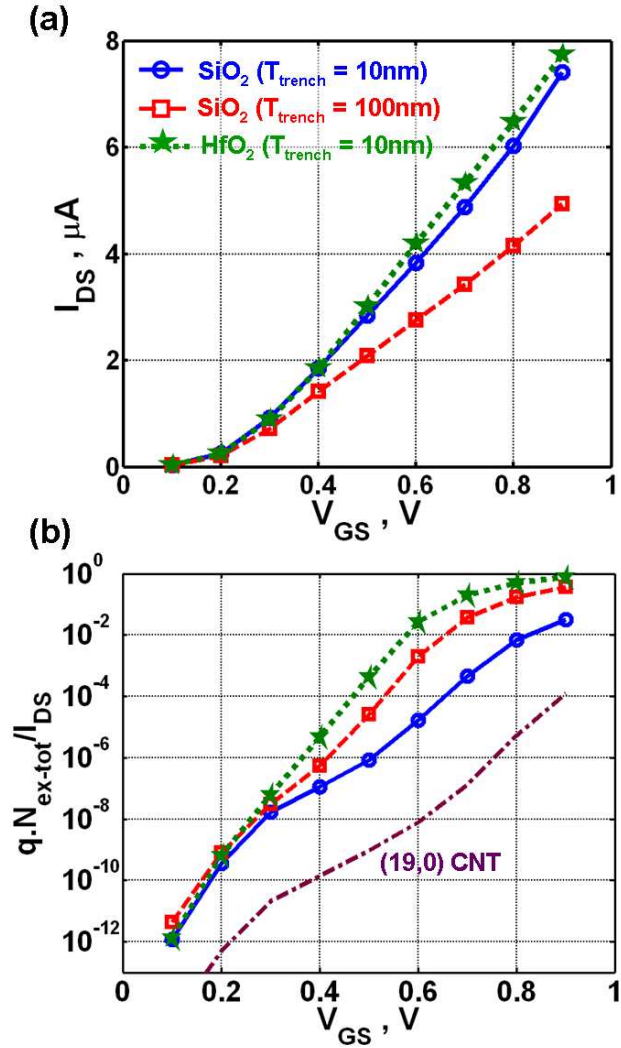

Fig. 5. (a) $I_{DS}$-$V_{GS}$ comparison at $V_{DS} = 0.7$V for a (25,0) CNT. Device parameters used here are similar to that in Fig. 2(a) except for $T_{trench}$ and the substrate oxide. $T_{OX} = 10$nm, $\varepsilon_{OX} = 3.9$ (SiO$_2$) and $\varepsilon_{OX} = 16$ (HfO$_2$) have been used. A linear increase in $I_{DS}$ is observed above the threshold voltage. On-current is reduced for the deeper trench due to reduced gate coupling in that region. (b) Exciton generation efficiency at the trench-substrate junction, $\eta_{exc} = (qN_{ex-tot}/I_{DS})$ vs. $V_{GS}$ shows an exponential increase. It is seen that the generation efficiency increases for deeper trench depths. High-k substrate oxide shows an even larger improvement. Exciton generation for a (19,0) CNT device with SiO$_2$ substrate oxide and $T_{trench} = 10$nm (dash-dot curve) is compared to that of the (25,0) CNT devices (see legend of (a)).

Device optimization schemes are discussed next. Figure 5 compares two different trench depths, $T_{trench}$ = 10nm and 100nm (solid circles and dashed squares), both with $SiO_2$ substrate oxide. Figure 5(a) shows that the deeper trench delivers a smaller device current. This is due to the weaker gate control of the electronic bands inside the trench region for deeper trench depths. More importantly, however, Fig. 5(b) shows that the exciton generation efficiency is higher for the device with a deeper trench which would result in higher optical emission intensities. This is due to the larger potential drop at the trench-substrate junction for those devices leading to a hotter carrier distribution and higher exciton generation rates. It is, however, observed that the improvement of emission efficiency with increasing trench depth has a saturation type behavior due to increased electrostatic screening effects inside the trench region.[28] Therefore, performance improvements attainable with deeper trench geometries could be limited.[28]

Figure 5 also compares the effect of high-k substrate oxides on device characteristics, i.e. $SiO_2$ (solid circles) vs. $HfO_2$ (dotted stars). It is seen that the current drive for the two structures is nearly equal since the device current is mainly determined by the modulation of the channel barrier inside the trench region; in this case, both devices have $T_{trench}$ = 10nm. On the other hand, the exciton generation efficiency for the latter device is significantly larger than that for $SiO_2$. This is due to the stronger gate coupling in the substrate region for the high-k oxide that leads to a larger potential drop at the trench-substrate junction (see Fig. 2(a)). This results in a hotter carrier distribution in that region and enhancement of exciton emission efficiency. Therefore, high-k substrate oxides could lead to significant improvement of optical emission intensity. It is noted that the emission efficiency could be further enhanced by using a deeper trench along with high-k oxides.

In conclusion, we have presented detailed numerical simulations of the partially suspended CNT optoelectronic device reported in Ref. 8. As observed in the experiment, our simulations show localized exciton generation by hot carriers, and an exponential increase in emission rate vs. linear increase in $I_{DS}$- $V_{GS}$. The emission efficiency is predicted to improve by deeper trench geometries. High-k substrate oxides could lead to even greater performance enhancements.

*Acknowledgement* - S. O. Koswatta thanks Dr. Sayed Hasan (Texas Instruments) and Prof. M. A. Alam (Purdue University) for fruitful discussions, and the Intel Foundation for Ph.D. Fellowship support. Computational support was provided by the NSF Network for Computational Nanotechnology (NCN).

**Supporting Information Available:** Description of BTE boundary conditions used in treating the Schottky barrier tunneling. This material is available free of charge via the Internet at http://pubs.acs.org.